\documentclass[12pt]{iopart}
\usepackage{iopams}
\usepackage{graphicx}
\usepackage{epsf}

\begin{document}

\sloppy

\jl{2}
\title[ionization of atomic hydrogen and He$^{+}$ by slow antiproton]
{\bf Ionization of atomic hydrogen and He$^{+}$ by slow antiprotons}

\author{\bf S Sahoo S C Mukherjee and H R J Walters}

\address{Department of Applied Mathematics and Theoretical Physics, 
Queen's University, Belfast BT7 1NN, UK}
\ead{s.sahoo@am.qub.ac.uk,j.walters@am.qub.ac.uk}

\begin{abstract}
We study the ionization process involving antiproton $(\bar p)$ and hydrogen in the energy range between 0.1 keV to 500 keV, using 
single center close coupling approximation. We construct the scattering wave function
using B-spline bases. The results obtained for ionization of atomic
hydrogen are compared with other existing theoretical calculations as well
as with the available experimental data. The present results are found
to be encouraging. We also employed this method to study the ionization
of He$^{+}$ in the energy range between 1 and 500 keV. On comparision, the present results are found to interpret
well the cross section values calculated using other theories.
\end{abstract}

\submitto{\JPB}

\maketitle
\section{Introduction}\label{sec:intro}
 The recent experimental research using slow antiprotons
$(\bar p)$ has been in progress. In near future it will measure the cross
sections in the low energy region and will provide a strong challange to
theory in order to predict  accurate cross sections in the energy
range the experiment is concerned. The collisions of $\bar p$ with
atomic hydrogen can be  considered as a fundamental process and is relevent in
many applied areas of physics. For proton impact, the final state can be
a superposition of elastic scattering, excitation, ionization and charge
exchange. However, for antiproton impact the charge transfer channel is
absent. Inspite of this simplicity, this process needs a careful
treatmnent easpecially in the case of slow $\bar p$ projectile. In low
energy heavy particle collision it is not easy to single out the
dominant channel, because many inelastic channels strongly couple with
one another open up, exchanging flux and phase in complicated
manner. Thus without inclusion of important channels, an accurate
determination of cross sections is not possible. In case of ionization
it is particularly important to describe the continuum part of the wave
function with utmost care in order to achieve accurate results.             

Recently, there have been a large number of studies for $\bar p$- H
system using various theoretical approachs. However, most of the close
coupling calculations are concentreted on single center
expansion method. It has been realized that accurate cross sections can be
calculated if a single-centred basis includes states with high angular
momenta. Because states associated with high angular momenta are capable
of describing the two center nature of the collision processes  and is
particularly suitable for $\bar p$ scattering (Hall \etal 1994, 1996,
Wherman \etal 1996). However, it has been reported by Toshima (2001)
that below 1 keV the one center pseudostate expansion method underestimates
the cross sections due to inability to represent the expanding
distribution of ionized electrons. For ${\bar p}$ projectile, most of
the calulations performed are single center close coupling methods, based on semiclassical impact parameter treatment where the
scattering wavefunctions are expanded around the target nucleus using
suitable bases (Schiwietz 1990, Hall \etal 1996, Igarashi \etal 2000, Azuma \etal 2002). Pons (1996) proposed a new momocentric
close coupling expansion in terms of spherical Bessel functions confined
in a finite box in the study of $\bar p$ - H ionization. Other methods
include direct solution of Schrodinger equation: Wells \etal (1996)
solved the Schrodinger equation directly on three dimensional lattice
without using expansion of basis set. Similarly Tong \etal (2001)
solved the Schrodinger equation taking a semiclassical approximation
for nuclear motion and the time evolution of electron wave function is
propagated by split-operator method with generalized pseudospectral
method in the energy representation. Sakimoto (2000) solved the time
dependent Schrodinger equation directly using a discrete variable
representation technique. For radial coordinates, he constructed the
numerical mesh from generalized laguerre quadrature points.    

In this article we make use of B-spline bases for the construction of
scattering wave function. B-spline has been widely used in
atomic physics (Martrin 1999) particularly due to its ability to describe the
continuum channels more accurately in
comparision to other conventional methods (Azuma \etal 2002). We give particular interest
to study the ionization of He$^{+}$ under $\bar
p $ impact. For the collision of $\bar p$ with hydrogenic ions such as
He$^{+}$,a number of calculations have been performed. Schultz \etal
(1996a) used four different methods to calculate the cross sections:
very large scale numerical solution of time-dependent Schrodinger
equation (TDSE), hidden crossing theory (HC), classical trajectory Monte
Carlo (CTMC), and continuum distorted eikonal initial state
(CDE-EIS). TDSE calculations which are assumed to be the most accurate in
the low energy region are found closer to HC results at low
energies. This calculation also follows 
CTMC results at intermediate energies and CDW-EIS results at high
energies. However, the TDSE cross sections are found to be about four
times larger than those calculated by Janev \etal (1995). A
discussion about this disagreements can be seen in the article by
Krstic \etal (1996). Wherman \etal (1996) used a large single centred Hibert basis sets to study the ionization of
He$^{+}$ by antiproton impact. They found that their results are in
good agreement with TDSE results, differeing by 6-13 \% and the results
obtained by Janev \etal (1995) were smaller by a factor of four.
Kirchner \etal (1999) used basis generator method (BGM) for $\bar p$ -
He$^{+}$ ionization. 
In case of $\bar p$ - H system, there is good convergency of results
among various theoretical approaches. However, the experimental data in
the low energy range is awiated. For the case of $\bar p$ - He$^{+}$ ionization
there is no experimental data available and it is necessary to
investigate this system in detail using different approachs and compare
the results with other theories. We study the ionization process of
hydrogenic ions under slow $\bar p$ projectile impact using  single center
expansion of scattering wave function in terms of B-spline basis sets. The
detailed description of present theory is presented in section
II. Atomic units ($a.u.$) are are used throughout unless otherwise stated.

\section{Theory}\label{sec:theo}

We use impact parameter approximation where the internuclear motion is
treated classically as $\bf{R} =\bf{b}+\bf{v}$t, with $\bf{b}$ the impact
parameter, $\bf{v}$ the impact velocity and $t$ the time and the electronic
motion is subjected to quantum mechanical laws. The electronic motion
can be described by the solution of time dependent Schrodinger equation 
\begin{equation}\label{eq:SE}
(H_{0}+V_{int}-i\frac{\partial}{{\partial}t})\Psi(\bf{r,R})=0
\end{equation}
where $\bf{r}$ is the position vector of the electron with respect to
proton. The atomic Hamiltonian is defined as 

\begin{equation}\label{eq:H0}
H_{0}=-\frac{1}{2}{\bigtriangledown^{2}}-\frac{Z_{T}}{r_{T}}
\end{equation}

where $Z_{T}$ is the nuclear charge of the target and $V_{int}$ is the
time depenent interaction between the projectile and target
electron. The interaction between $\bar p$ - hydrogenic ions is given by 

\begin{equation}\label{eq:Vint}
V_{int}=\frac{1}{\vert{\bf{r-R}}\vert}
\end{equation} 

The total wave function is expanded as
 
\begin{equation}\label{eq:w1}
\Psi_{nlm}($\bf{r}$)=\sum_{nlm}a_{nlm}(t)\phi_{nlm}($\bf{r}$)exp(-i\varepsilon_{nlm}t),
\end{equation}

\begin{equation}\label{eq:f2}
\phi_{nlm}($\bf{r}$,t)=F_{nl}(r)[(-1)^{m}Y_{lm}($\bf{r}$)+Y_{l-m}($\bf{r}$)]/\sqrt2(1+\delta_{m,0}).
\end{equation}

The radial part of the wave function is further expanded as 
\begin{equation}\label{eq:f3}
F_{nl}(r)=\sum_{i}c_{ni}\frac{B_{\imath}^{k}(r)}{r}
\end{equation}
where $B_{\imath}^{k}$(r) is the k-th order B-spline functions. The entire space of
the electron sphere is confined with radius $r=r_{max}$. The interval
[0, r$_{max}$] is then devided into segments. The end points of these segments
are given by the knot sequence $t_{\imath}$, $\imath$=1,2,....n+k. B-splines are
piecewise polynomials of order k defined recursively on this knot
sequence via the formule: 

\begin{eqnarray}
B_{i}^{1}(r)= \left \{\begin{array}{ll}
1& t_i\leq r < t_{i+1}\\
0 & \mbox{otherwise}\,\,\,\,,
\end{array}\right.
\end{eqnarray}

and 

\begin{eqnarray}\label{B1}
B_{i}^{k}(r)=\frac{r-t_i}{t_{i+k-1}-t_i}B_{i}^{k-1}(r)+ \frac{t_{i+k}-r}{t_{i+k}-t_{i+k}}B_{i+1}^{k-1}(r).
\end{eqnarray}

Each B-spline is a piecewise polynomial of degree $k$-1 inside the
interval $t_{\imath} \le{r} <{t_{\imath+1}}$ and zero outside the
interval.The piecewise nature of B-splines are ideally suited to
represent atomic wave functions. We chose an exponential knot sequence
so as to model the exponential behaviour of the wavefunctions. 
For the radial function
to satisfies the boundary condition that $F_{nl}(0)=0,$ and $F_{nl}(r)=0$
at  $r=r_{max}$, we omit the first and last B-splines respectively. The
coefficients $c_{ni}$ of B-spline are determined by diagonilizing the
atomic Hamiltonian $H_{0}$, 

\begin{equation}\label{eq:f4}
<\phi_{n'l'm'}\vert{H_{0}}\vert\phi_{nlm}>=\varepsilon_{nl}\delta_{nn'}\delta_{ll'}\delta_{mm'}
\end{equation}

The eigen energies obtained for lowest eight eigen states are found to
be closer to the exact ones.

By substituting equation (4) into the Schrodinger equation (1) we have
coupled equations with respect to the expansion coefficients $a_{n'l'm'}(t)$,

\begin{equation}\label{eq:f5}
i\frac{d}{dt}a_{n'l'm'}(t)=\sum_{nlm}exp[i(\varepsilon_{n'l'm'}-\varepsilon_{nlm})t]<\phi_{n'l'm'}\vert{V_{int}}\vert\phi_{nlm}>a_{nlm}(t).
\end{equation}

The above coupled equations are solved with the initial condition
$a_{nlm}(\bf{b},-\infty)$=$\delta_{n'l'm',1s}$.

The sum of the probabilities $P_{nlm}(b)$ = $\vert{a_{nlm}}(b)\vert^{2}$ over eigen states
with positive energies gives the ionization probability for a particular
impact parameter. The ionization cross section can be obtained as 
          
\begin{equation}\label{eq:cross}
\sigma= 2\pi \int_0^{\infty}P_{nlm}(b)bdb.
\end{equation}

\section{Results and Discussions}\label{sec:result}

We solved the Schrodinger equation for $\bar p$ colliding with hydrogen
and hydrogenic ions.  Calculations are performed with 45 radial
functions obtained from 8th order B-splines defined in the interval 0 to
$r_{max}$=200 a.u. The maximum orbital angular momentum $l_{max}$
used in the calculation is 8. Since a single centred expansion calculation requires the retaintion
of much higher values of angular momentum for producing well converged results. By
taking all the degeneracies for magnetic quantum number $m$, we solved
the coupled differential equation (10) with 2025($m\geq0$) number of basis sets. We
integrated equation (10) in the interval $vt=-30$ $ a.u.$ to $vt=50
$ $a.u.$. we considered the motion of the projectile along z axis and x-z
is the collision plane.   
 
Figure 1 dispalys the total cross sections for $\bar p$-H
ionization. For comparison we also displayed the results obtained from
other theoretical approaches. 
\begin{figure}[h]
\includegraphics[width=\textwidth]{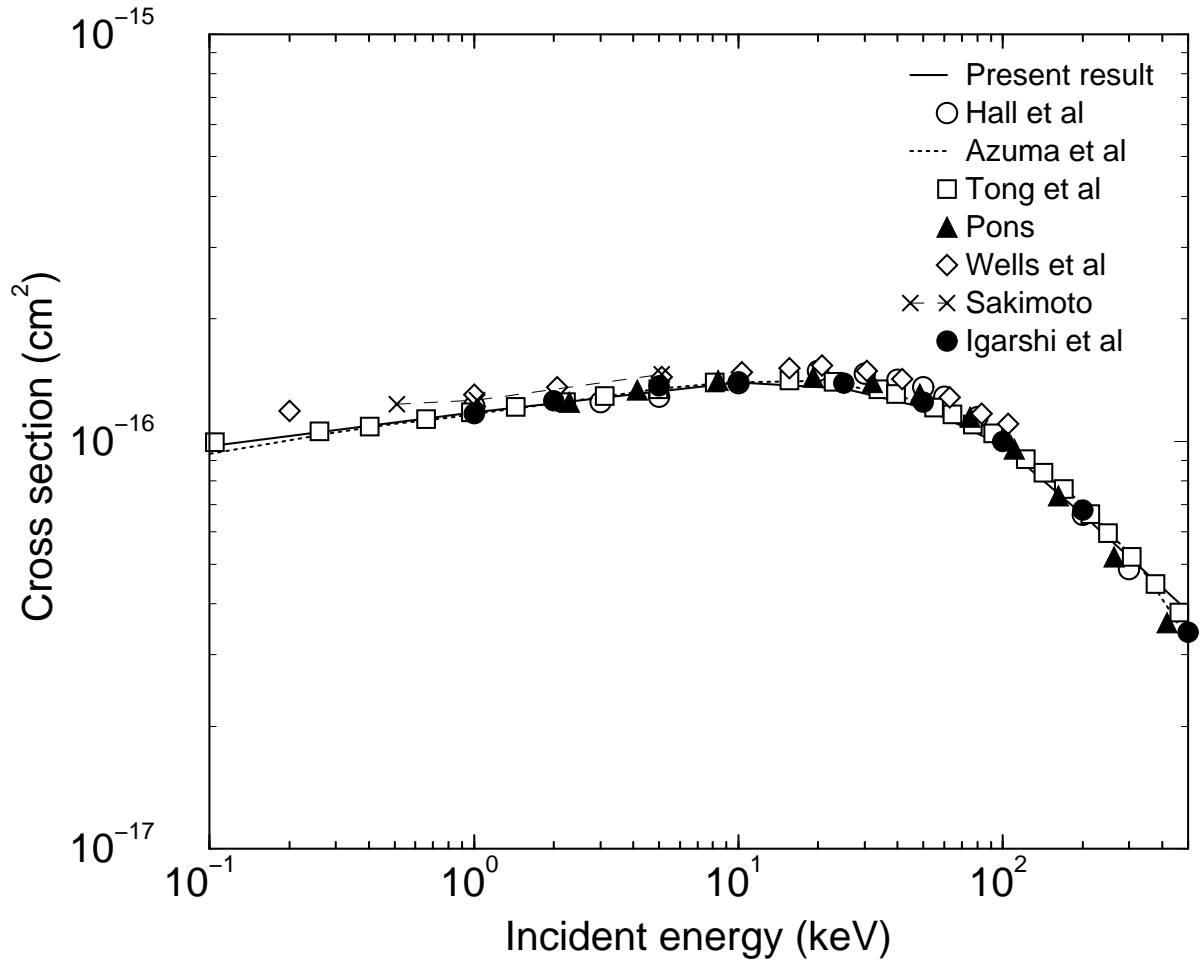}
\caption{Total ionization cross sections of H under $\bar p$ impact}
\label{fig:1}
\end{figure}

The present calculated ionization cross sections are found to be in good
agreement with the results of Tong \etal (2001)  throughout the energy
range considered. This calculation has been carried out with straight line
trajectory. It may be mentioned that these authors also
performed a calculation using curved trajectory which is not presented
here.  The results of one center Hilbert space calculation Hall \etal are found to be a little higher than the present calculated values
in the energy range between 20-100 kev impact energies. However, the
once center calculation of Igarshi \etal which uses Sturmian basis
is found to be in resonably good
agreement with the present values. The results of Sakimoto (2000) who
used Laguerre meshes and the TDSE results of Wells \etal are a little
higher than our calculated values. Wells \etal used  a numeriucal
solution of three dimensional Cartesian co-ordinate grids. The results
of direct solution is always larger than the other calculated
values. They mentioned that consideration of only $n\leq 3$ bound
channels is insufficient and the estimated cross sections would
overestimate. The calculation of Pons which makes use of the spherical
Bessel functions to describe continuum channels are consistent with the
present calculation. The recent calculation of Azuma \etal who used the B-spline
bases similar to the present one is found be in better agreement except
around 0.1 keV impact energies. At this incident energy the present
value slightly overestimates the calculation of Azuma \etal.

\begin{figure}[h]
\epsfxsize=10cm
\includegraphics[width=\textwidth]{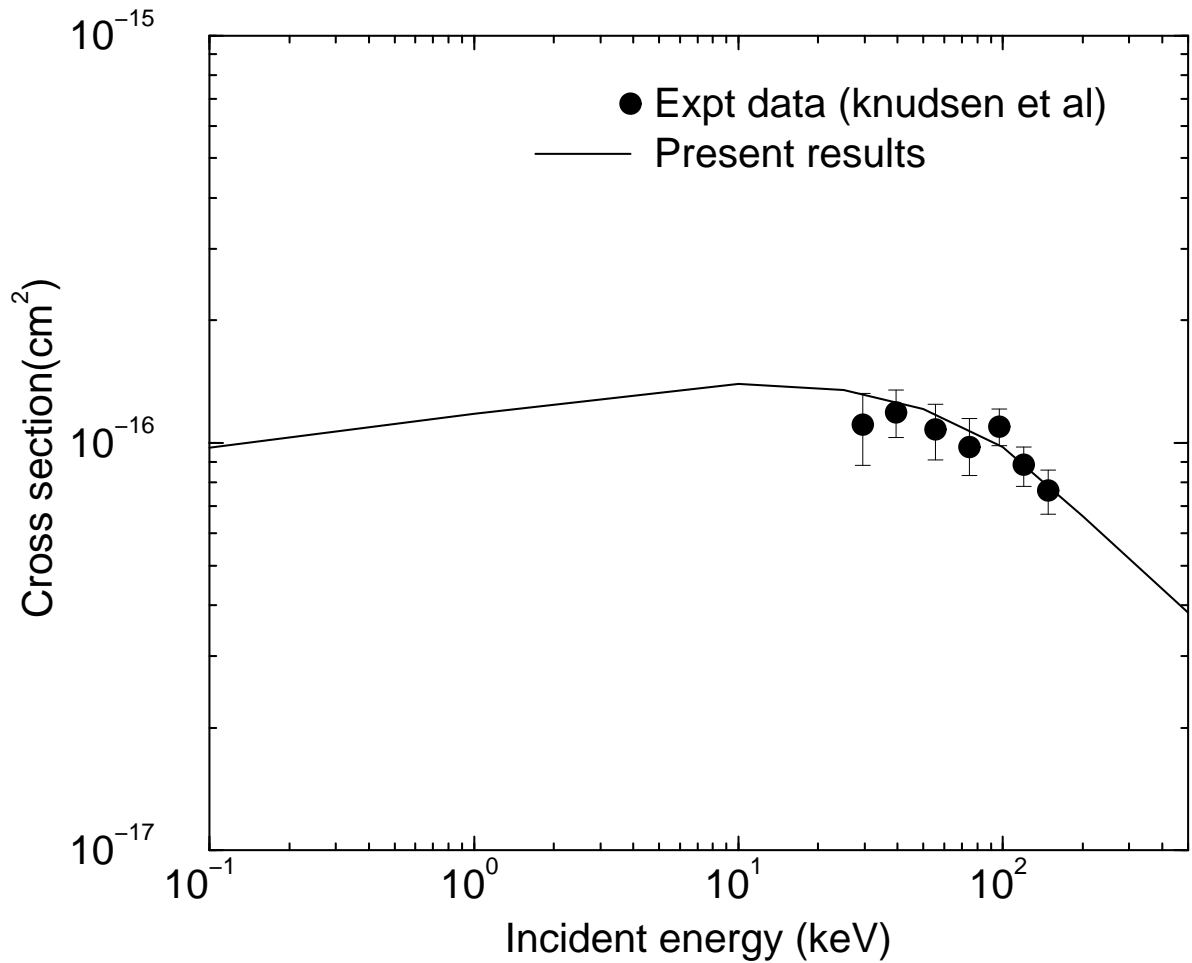}
\caption{Total ionization cross sections in $\bar p$ - H collisions,
solid line : present results, closed circle : expt.data (Knudsen \etal (1995)}
\label{fig:2}
\end{figure}

In figure 2 we compare the present calculated results with available
experimental measurements of Knudsen \etal (1995). There  is a good
agreement between the present results and the experimental data over the
whole energy range considered.

It clear from Figs. 1 that all the
results for total ionization calculated using different approaches show
reseonably good agreement in both qualitative and quantitative
measures. All the theoretical values including the present one (Fig. 2) are in
good agreement with the experiment of Knudsen \etal. It will be interesting if the experiment measures the cross
section data down to 1 keV energy range. We hope these will be available soon.
\begin{figure}[h]
\epsfxsize=10cm
\includegraphics[width=\textwidth]{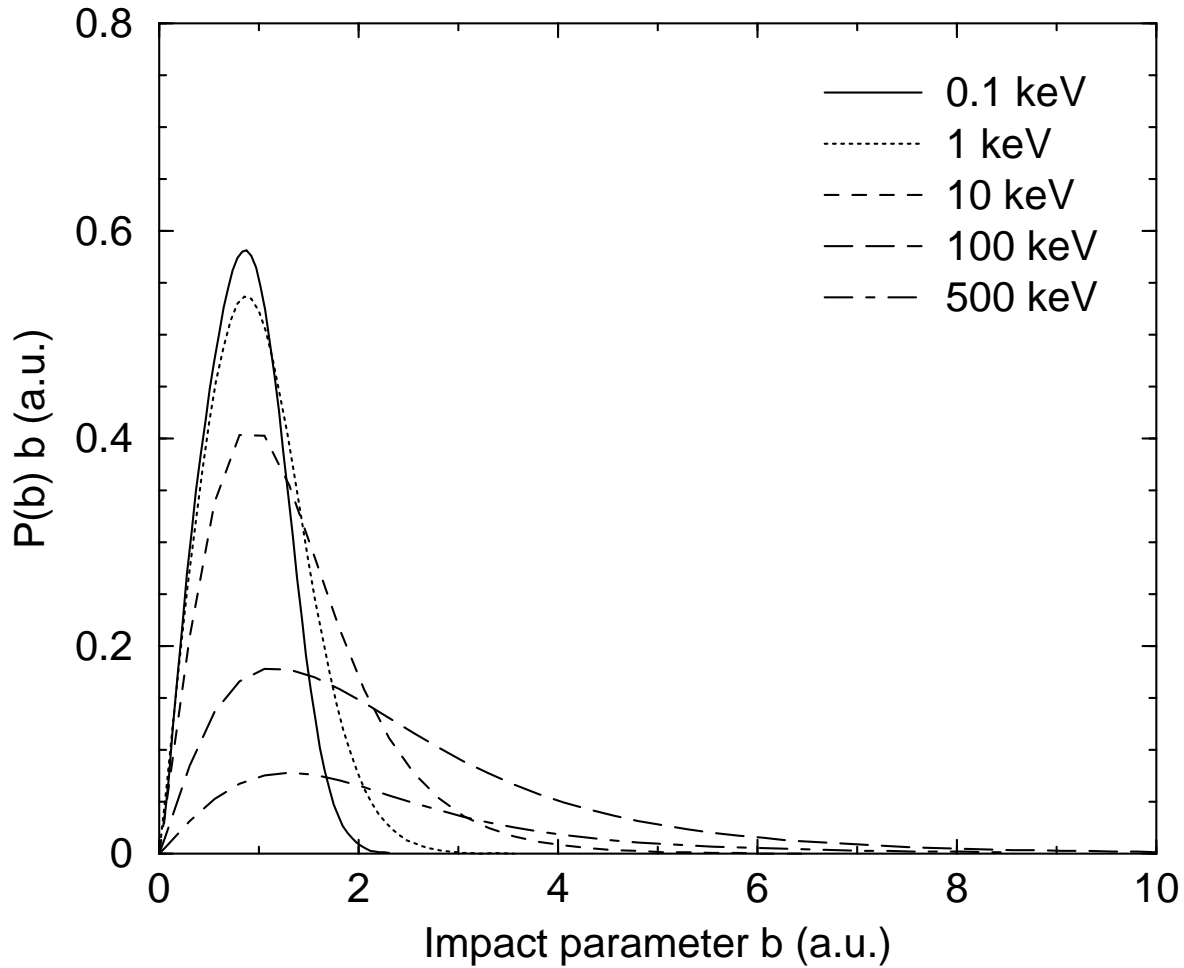}
\caption{Ionization probabilities in $\bar p$ - H collisions as a
function of impact parameter $b$ at various incident energies}
\label{fig:2}
\end{figure}
In figure 3 we plotted $b P(b)$ as a function of impact parameter $b$
for several incident energies. 
It may be observed from the figure that the probability for high impact energy
shows long tail and it dissapears as the collision energy
decreases. The paek values also shifts to the lower impact parameter as
the collision energy decreases. 

In Fig. 4 we display the results of our single center B-spline basis set
calculation for total ionzation cross section of He$^{+}$ by ${\bar
p}$ impact for a wide energy range from 1 keV to 500 keV. 

\begin{figure}[ht]
\includegraphics[width=\textwidth]{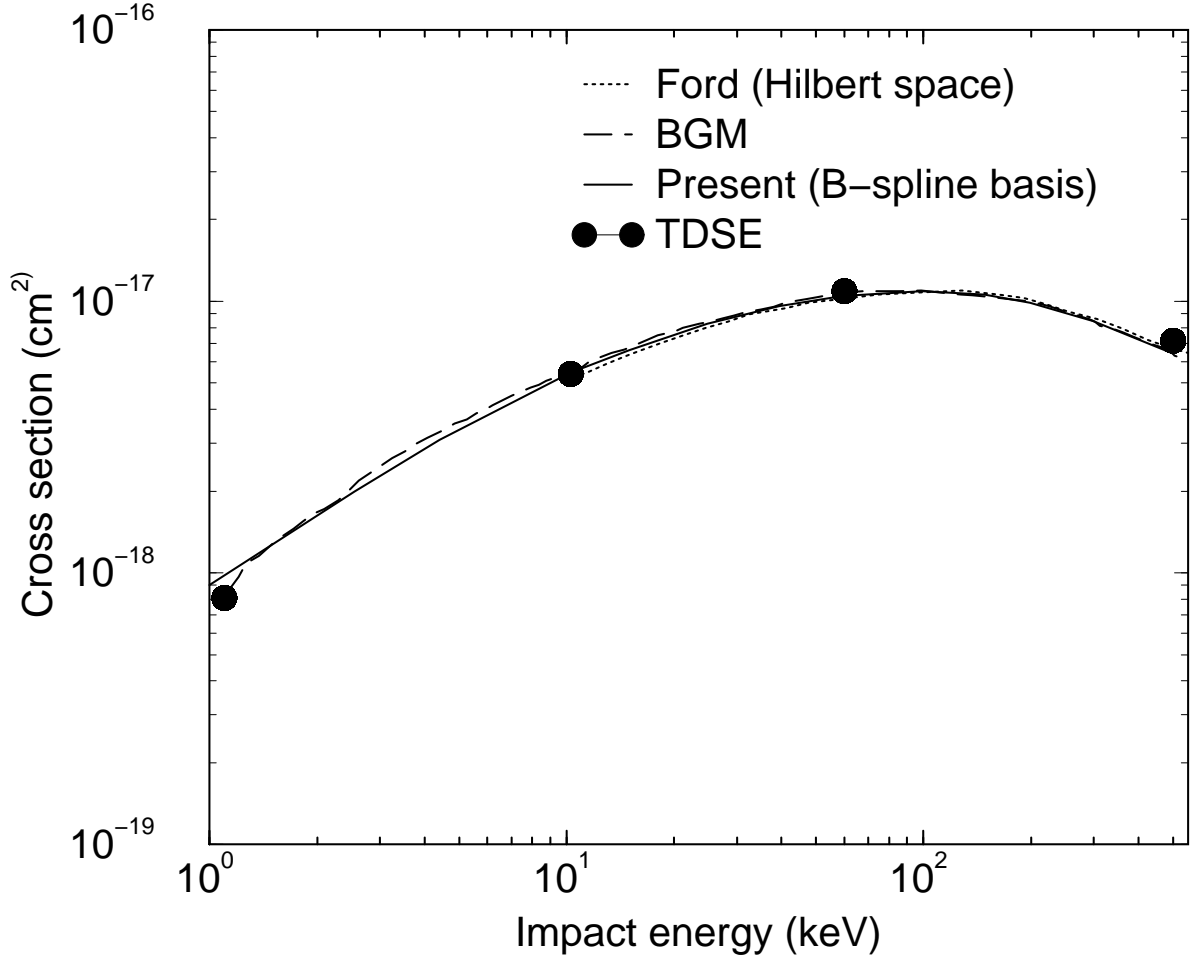}
\caption{Total ionization cross section of He$^{+}$(1s), solid line: present results, solid circle: LTDSE (Schultz \etal
1997), dashed line: SCE (Ford A L, Private communication), long dashed
line: BGM (Kirchner \etal 1999)}
\label{fig:3}
\end{figure}

For comparasion we also show in the figure, the results obtained
by Lattice Schrodinger-equation approach (LTDSE) (Schultz \etal 1997) and
Single center results (Ford \etal (private communication), Wherman
\etal (1996) and references therein). Also included in the figure are the
results of Kirchner \etal who used Basis Generator Method
(BGM). This method deals with the construction of a basis that
dynamically adapts to the collision process considered in order to
follow the propagation and to cover the one dimensional subspace defined
by the solution of the time dependent Schrodinger equation(TDSE).  It
may be seen in the figure that the present results are found to be in
good agreement with the calculation of Ford \etal who employed a
single centred Hilbert basis set. However, both LTDSE and BGM results
slightly 
overestimate the present cross sections. It has been mentioned in the
paper of Schultz \etal that this overestimation in comparision to the
results of Ford \etal is about 10\%. They
reported that this may due to the fact that it is likely that the excitation of
He$^{+}$ to higher $n$ values probably $n\geq4$,  which has been incorrectly
treated as ionization in LTDSE grid. They finally concluded that the
treatment of excitation to $n\geq4$ would be important in their
method. Additionally other factors such as grid spacing would also needs
to be carefully examined in order to calculate the ionization result
beyond an accuracy of 10\%. 
\begin{figure}[h]
\includegraphics[width=\textwidth]{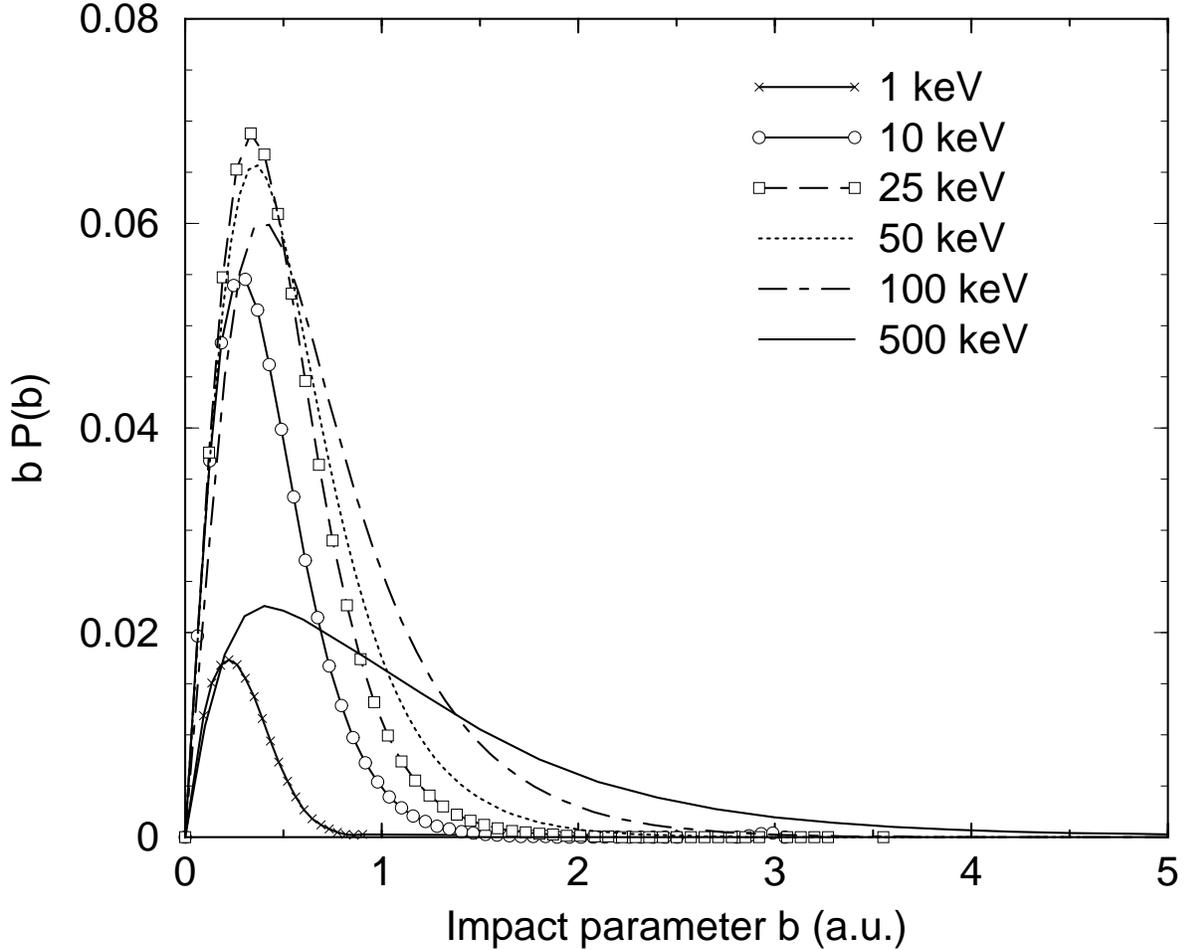}
\caption{Ionization probabilities in $\bar p$ - He$^{+}$(1s) collisions as a
function of impact parameter $b$ at various incident energies}
\label{fig:4}
\end{figure}

 To support the present calculation, we displayed in Fig. 5 the variation of
 ionization probability with the impact parameter for various collision
energies. It may be noted that $b~P(b)$ as function of $b$ 
shows long tail for higher impact energies which is well experienced
in case of 500 keV impact energies. However,  for all collision energies
 peaks around b = 0.3 $a.u.$
 are observed. We also derived
the dynamic ionization probability at this impact parameter (0.3 $a.u.$)
 when the two nuclei are separated from each other at some
 distance. 

\begin{figure}[h]
\includegraphics[width=\textwidth]{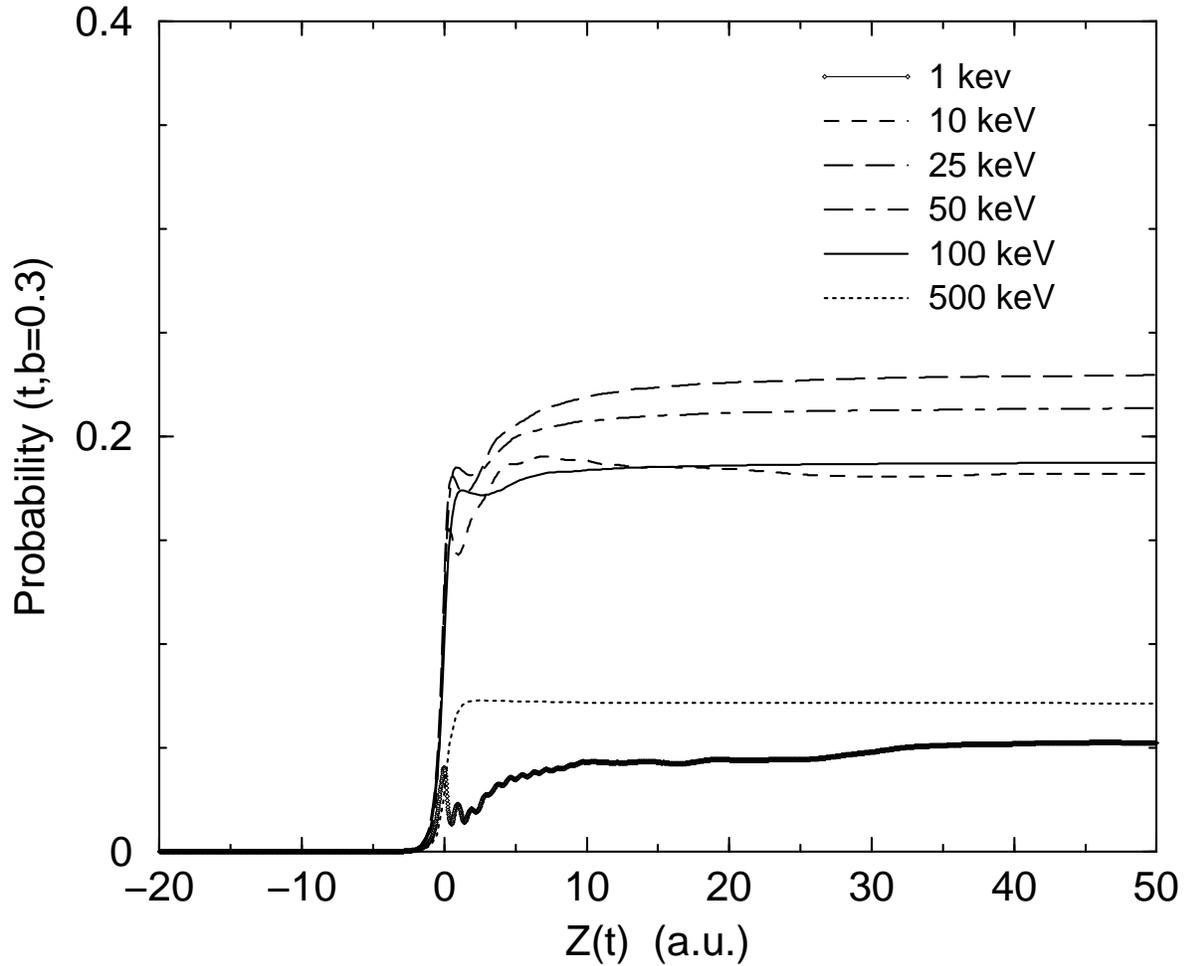}
\caption{The dynamic probabilities in $\bar p$ - He$^{+}$(1s) ionization at various
incident energies for a particular impact parameter $b$ = 0.3 a.u.}
\label{fig:5}
\end{figure}
This is shown in Fig. 6. It is clear from the figure that the
 ionization probabilities saturate around $z(vt)$ = 10  
$a.u.$ for all impact energies except 500 keV where the saturation
 starts early around $z(vt)=3 a.u.$. The probability shows a rapid
growth between -3 and 3 $a.u.$ and then saturatation starts.  Therefore we allowed
 sufficient time for the probability to become completely stable. The
 same type of situation has been shown by Tong \etal (2001) who used curved
 trajectory for  $\bar
 p$ - H ionization. They have reported that for high collision energies (above
 1 keV) the probability saturates z(vt) above 10 $a.u.$ with a rapid
 increase from Z(vt) = -5 $a.u.$. For collision velocities (below 1
 keV), they found a slow increase of the probability. This time delay
 can be termed as post-collisional interaction. Pons (2000) indicated
 that that due to slow antiproton motion the projectile pushes away the
 ejected electron even when the prjectile is going farther from the
 target. Afterall in the present case Fig. 5 helps for a convergence
 check.  

It may be worth to mention that in the case
of $\bar p$ - H ionization, where in the limit of small internuclear
distance, the electron experiences a dipole like potential bound by two nuclei (Krstic \etal 1996) and there exists a critical value of the dipole
stength below which no bound state can be supported. It corresponds to
the internuclear distance known as Fermi-Teller radius (Fermi and Teller
1947) at which the eigen energies of the ground state merge with the continuum. However, in case
of an asymmetric dipole as in $\bar p$ - He$^{+}$ case, Krstic \etal
reported that  the electronic eigen states donot merge with the
continuum and hence the ionization cross sections are expected to show
an exponential decrease for small collision velocities. These situations
are clearly evient in Fig. 1 and Fig. 2.

\section{Conclusions}\label{}

  The results obtained for $\bar p$ - H are found to be in good agreement
  with the other calculated values as well as the available experimental
  data. However, the experimental results are still awaited in low
  energy range. For the case
  of He$^{+}$ target, all the theoretical calculations includding the
  present one show good agreement within a few percent of
  accuracy. Specifically the present results and the results of Ford
  show a good convergency. Our B-spline basis results confirms the
  single center Hilbert space calculation of Ford. There is no measured
  values for this system. It would be interesting to have  more calculations for slow $\bar p$ projctile collisiding with
  He$^{+}$.

\section{Acknowledgement}

 One of the authour SS is thankful to G. Gribakin for his valuable
 discussions. We also greatly acknowledge the financial support from
 Queen's University Belfast under IRCEP programme.   
  
\section*{References}

\begin{harvard}

\item[] Hall K A, Reading J F and Ford A L 1996
{\em J. Phys. B: At. Mol. Opt. Phys}. {\bf 29} 6123-6131; 1994 {\bf
27} 5257-5270

\item[] Wherman L A, Ford A L and Reading J F 1996 {\em J. Phys. B: At. Mol. Opt. Phys}. {\bf 29} 5831-5842

\item[] Toshima N 2001 {\em Phys. Rev}. A {\bf 64} 024701

\item[] Schiwietz {\em et al.} 1996 {\it J. Phys. B:
At. Mol. Opt. Phys}. {\bf 29} 307

\item[] Igarashi A, Nakazaki S and Ohsaki 2000 {\em
Phys. Rev}. A {\bf 61} 062712

\item[] Azuma Z, Toshima N and Hino K 2002 {\em Phys. Rev.} A
{\bf 64} 062704

\item[] Pons B 2000 {\em Phys. Rev. Lett. {\bf 84} 4569};{\em
Phys. Rev}. A {\bf 63} 012704

\item[] Wells J C, Schultz D R, P. Gavras P and Pindzola 1996
{\em Phys. Rev}. A {\bf 54} 593 

\item[] Xiao-Min T, Watanabe T, Kato D and Ohtani S 2001 {\em
Phys. Rev}. A {\bf 64} 022711 

\item[] Sakimoto K 2000 {\em
J. Phys. B:At. Mol. Opt. Phys.}{\bf33} 5165

\item[] Martrin F 1999 {\em J. Phys. B: At. Mol. Opt. Phys}. {\bf 32}
R197

\item[] Schultz D R, Krstic P S, Reinhold C O anf Wells
J C 1996a {\em Phys. Rev. Lett}. {\bf 76} 2882-5

\item[] Janev R K, Solov'ev E A and Jakimovski D 1995 {\em
J. Phys. B: At. Mol. Opt. Phys}.{\bf 28} L615-20

\item[] Krstic P S, Schultz D R and Janev R. K. 1996
{\em J. Phys. B: At. Mol. Opt. Phys.} {\bf 29} 1941-68

\item[] Kirchner T, Ludde H J, Kroneisen O. J, and
R. M. Dreizler 1999 {\em Nucl. Instrum. Meth.} B {\bf 154} 46

\item[] Schultz D R, Wells J C, Krstic and Reinhold C O
1997 {\em Phys. Rev}. A {\bf 56} 3710 

\item[] Ford A L (private communication). The results are
taken from the paper of Schultz {\etal} 1997

\item[] Fermi E and Teller E 1947 {\em Phys. Rev}. {\bf72} 399
\item[] Knudsen H, Mikkelsen U, Paludan K, Kirsebom K, Moller S P,
Uggerhoj E, Slevin J, Charlton M and Morenzoni E 1995{\em
Phys. Rev. Lett.}  {\bf 74} 4627
 
\end{harvard}




\end{document}